\documentclass[12pt]{article}
\usepackage{a4wide}
\usepackage{amssymb}
\usepackage{graphicx}
\usepackage{subfigure}
\usepackage{epsfig}
\usepackage[tbtags]{amsmath}
\usepackage{exscale}
\begin{document}

\begin{center}
{\Large Modified constraint algebra in loop quantum gravity and spacetime interpretation} \\
\vspace{1.5em}
Rakesh Tibrewala \footnote{e-mail address: {\tt rtibs@iisertvm.ac.in}}
\\
\vspace{1em}
Indian Institute of Science Education and Research,
CET Campus, Trivandrum 695016, India
\end{center}

\newcommand{\lP}{\ell_{\mathrm P}}

\newcommand{\md}{{\mathrm{d}}}
\newcommand{\sgn}{\mathop{\mathrm{sgn}}}

\newcommand*{\R}{{\mathbb R}}
\newcommand*{\N}{{\mathbb N}}
\newcommand*{\Z}{{\mathbb Z}}
\newcommand*{\Q}{{\mathbb Q}}
\newcommand*{\C}{{\mathbb C}}

\newcommand{\Ef}{E^\varphi}
\newcommand{\Kf}{K_\varphi}

\newcommand{\kp}{\ensuremath{K_\varphi}}
\newcommand{\kx}{\ensuremath{K_x}}
\newcommand{\ex}{\ensuremath{E^x}}
\newcommand{\ep}{\ensuremath{E^\varphi}}
\newcommand{\gp}{\ensuremath{\Gamma_\varphi}}
\newcommand{\mrl}{\ensuremath{\sqrt{2Mx}}}
\newcommand{\dkp}{\ensuremath{\,\delta\!K_\varphi}}
\newcommand{\dkx}{\ensuremath{\,\delta\!K_x}}
\newcommand{\dep}{\ensuremath{\,\delta\!E^\varphi}}
\newcommand{\dex}{\ensuremath{\,\delta\!E^x}}
\newcommand{\dn}{\ensuremath{\,\delta\!N}}
\newcommand{\dnx}{\ensuremath{\,\delta\!N^x}}
\newcommand{\dnt}{\ensuremath{\,\delta_2 N}}
\newcommand{\dgop}{\ensuremath{\delta_1 \Gamma_\varphi}}
\newcommand{\dgtp}{\ensuremath{\delta_2 \Gamma_\varphi}}
\newcommand{\dkop}{\ensuremath{\delta_1 K_\varphi}}
\newcommand{\dktp}{\ensuremath{\delta_2 K_\varphi}}
\newcommand{\dktx}{\ensuremath{\delta_2 K_x}}
\newcommand{\detp}{\ensuremath{\delta_2 E^\varphi}}
\newcommand{\detx}{\ensuremath{\delta_2 E^x}}
\newcommand{\dntx}{\ensuremath{\,\delta_2\!N^x}}
\newcommand{\be}{\begin{equation}}
\newcommand{\ee}{\end{equation}}
\newcommand{\dif}{\mathrm{d}}
\newcommand{\dkpp}{\ensuremath{\,\delta\!K'_\varphi}}
\newcommand{\dkxp}{\ensuremath{\,\delta\!K'_x}}
\newcommand{\depp}{\ensuremath{\,\delta\!E^{\varphi'}}}
\newcommand{\dexp}{\ensuremath{\,\delta\!E^{x'}}}
\newcommand{\dnxp}{\ensuremath{\,\delta\!N^{x'}}}
\newcommand{\dktpp}{\ensuremath{\delta_2 K'_\varphi}}
\newcommand{\dktxp}{\ensuremath{\delta_2 K'_x}}
\newcommand{\detpp}{\ensuremath{\delta_2 E^{\varphi '}}}
\newcommand{\detxp}{\ensuremath{\delta_2 E^{x'}}}

\begin{abstract}
Classically the constraint algebra of general relativity, which generates gauge transformations, is equivalent to spacetime covariance. 
In LQG, inverse triad corrections lead to an effective Hamiltonian constraint which can lead to a modified constraint algebra. We show, using example of spherically symmetric spacetimes, that a modified constraint algebra does not correspond to spacetime coordinate transformation. In such a scenario the notion of black hole horizon, which is based on spacetime notions, also needs to be reconsidered. A possible modification to the classical trapping horizon condition leading to consistent results is suggested. In the case where the constraint algebra is not modified a spacetime picture is valid and one finds mass threshold for black holes and small corrections to Hawking temperature.
\end{abstract}

Quantum gravity continues to remain elusive despite numerous attempts taking various forms. Loop quantum gravity (LQG) - a non-perturbative, canonical approach to quantum gravity, is one such attempt \cite{Rov, ThomasRev, ALRev}. Underlying its construction is a discrete notion of spacetime, a quantum geometry replacing smooth geometry of general relativity. 
In the context of symmetry reduced cosmological models, where there are only finitely many degrees of freedom (similar to quantum mechanics), the framework has been quite successful leading to a resolution of the big bang singularity \cite{AbSinginLQC, APS1}. However, this success comes with an important question mark - whether the singularity resolution is an artifact of classical symmetry reduction? For this one needs to go beyond the homogeneity of cosmology  and probe inhomogeneous (in other words, field theoretic) models. Spherically symmetric models seem ideal for this purpose.

At present the dynamics of the theory is not fully understood. However, certain effects do not crucially depend on the details of the dynamics and can be used to make preliminary studies of inhomogeneous systems. One such effect which is easy to implement in the context of spherical symmetry is the inverse triad effect, coming from the quantization of inverse powers of the triad variables. This effect corrects the Hamiltonian of the theory leading to modified dynamics. More importantly, it can modify the constraint algebra of the theory. The (classical) constraint algebra serves a dual purpose - generating gauge transformations in the phase space and encoding spacetime covariance of the theory. A modification of the constraint algebra can break this correspondence. This in effect could mean that the classical spacetime concepts - the notion of black hole horizon for instance - no longer hold. We provide an explicit example to show that this indeed is the case. We suggest a modification of the classical horizon condition that leads to consistent results. In the case where the corrections do not modify the constraint algebra, classical spacetime notions are valid and lead to a mass threshold for black holes and small corrections to Hawking temperature (detailed calculations can be found in \cite{ModSpacetime}). 

\section{Classical theory and constraint algebra}

We begin with classical theory for spherical symmetry using Ashtekar variables which consist of $\mathfrak{su}(2)$-valued connection $A^{i}_{a}(x)=\Gamma^{i}_{a}(x)+\gamma K^{i}_{a}(x)$ and densitized triads $E^{a}_{i}(x)$. Here $\Gamma^{i}_{a}$ is the spin connection, $K^{i}_{a}$ is related to the extrinsic curvature, $\gamma$ is the Barbero-Immirzi parameter and $E^{a}_{i}$ determines the three metric $q_{ab}$. Hamiltonian formulation leads to the Hamiltonian and the diffeomorphism constraints (and an additional Gauss constraint in $\mathfrak{su}(2)$ variables). After solving the Gauss constraint, one is left with a set of two canonical pairs (see \cite{SphSymm, SphSymmHam} for details)
\be
\{\kx(x),\ex(y)\}=2G\delta(x,y)\quad\mbox{and}\quad\{\kp(x),\ep(y)\}=G\delta(x,y)\,.
\label{canonicalVars}
\ee

The spherically symmetric metric in terms of these variables is given by
\begin{equation} \label{SSMetric}
{\rm d}s^2=-N^2{\rm d}t^2+\frac{\Ef\,^2}{|E^x|}({\rm d}x+N^x{\rm
  d}t)^2+ |E^x|{\rm d}\Omega^2
\end{equation}
where  $N(t,x)$ is the lapse function and $N^x(t,x)$ is the shift vector. The Hamiltonian constraint $H[N]$ and the  diffeomorphism constraint $D[N^{x}]$ (including generic matter contributions) are
\be\label{C2M}
\mathbf{H}[N]=-\frac{1}{2G}\int \md x\, N|\ex|^{-\frac{1}{2}}\bigg[\kp^2\ep+2\kp\kx\ex+(1-\gp^2)\ep+2\gp'\ex-8\pi G\ep|\ex|\rho \bigg]\approx0
\ee
\be\label{D2M}
\mathbf{D}[N^x]=\frac{1}{2G}\int \md x\,N^x\bigg[2\ep\kp'-\kx E^{x'}-8\pi G \ep \sqrt{|\ex|}J_x \bigg]\approx0 \,,
\ee
where $\Gamma_{\phi}=-E^{x'}/2E^{\phi}$. The gravitational part of these satisfies the \emph{surface deformation algebra}
\begin{eqnarray}
\{H_{\rm grav}[N],D_{\rm grav}[N^x]\}&=&-H_{\rm grav}[N^xN']\,,
\label{classDefAgebraDiff}\\
\{H_{\rm grav}[N],H_{\rm grav}[M]\}&=&D_{\rm grav}[|E^x|(\Ef)^{-2}(NM'-MN')]\,.   \label{classDefAlgebra1}
\end{eqnarray} 
As discussed in \cite{Regained}, \eqref{classDefAgebraDiff} and \eqref{classDefAlgebra1} imply that dynamics takes place on spacelike hypersurfaces. 

\section{Inverse triad corrections and constraint algebra}
Next we consider LQG effect in the form of the inverse triad corrections. These arise when quantizing the inverses of triad operators appearing in the Hamiltonian. Since triads have a discrete spectrum containing zero, they do not have well defined inverses. Techniques exist \cite{QSDV} reproducing the inverse in the classical limit but implying corrections in the quantum domain. We will represent the effective inverse as $1/E\rightarrow\alpha(E)/E$, (see \cite{LTB} for derivation of $\alpha(E)$ and Fig.~\ref{alphaDelta} for its plot). 
The gravitational part of the Hamiltonian constraint in \eqref{C2M} thus becomes 
\begin{align}
H^Q_{\rm grav}[N]=-\frac{1}{2G}\int \md x\, N\bigg[\frac{\alpha\kp^2\ep}{|\ex|^{\frac{1}{2}}}+2\bar{\alpha}\kp\kx|\ex|^{\frac{1}{2}}+\alpha(1-\gp^2)\frac{\ep}{|\ex|^{\frac{1}{2}}}+2\bar{\alpha}\gp'|\ex|^\frac{1}{2} \bigg] \label{myEffectiveHamiltonian}
\end{align}
where, for generality, different powers of $E^{x}$ have been corrected by different functions $\alpha$ and $\bar{\alpha}$. 
(No corrections are included with spin connection $\Gamma_{\phi}$ since it does not lead to anomaly free algebra \cite{LTBII}). For the quantum corrected Hamiltonian, the constraint algebra turns out to be
\begin{eqnarray}
\{H_{\rm grav}^Q[N],D_{\rm grav}[N^x]\}&=&-H_{\rm grav}^Q[N^xN']\,,
\label{DefAgebraDiff}\\
\{H_{\rm grav}^Q[N],H_{\rm grav}^Q[M]\}&=&D_{\rm grav}[\bar{\alpha}^2|E^x|(\Ef)^{-2}(NM'-MN')]\,.   \label{DefAlgebra1}
\end{eqnarray} 
If $\bar{\alpha}=1$, the algebra is classical, implying that dynamics takes place on spacelike hypersurface. In general, $\bar{\alpha}\neq1$ and the algebra is modified and one would like to explore its implications with regard to the spacetime covariance property of the quantum theory. We will consider two cases - (i) $\bar{\alpha}=1$, in which case only the dynamics is modified but spacetime covariance properties are classical and (ii) $\bar{\alpha}=\alpha$ in which case both the dynamics and spacetime properties change (situations more general than these can be related to these two cases \cite{ModSpacetime}).

\section*{Case I: $\bar{\alpha}=1$} 
\begin{figure}
\begin{minipage}{0.5\linewidth}
\centering
\includegraphics[scale=0.75]{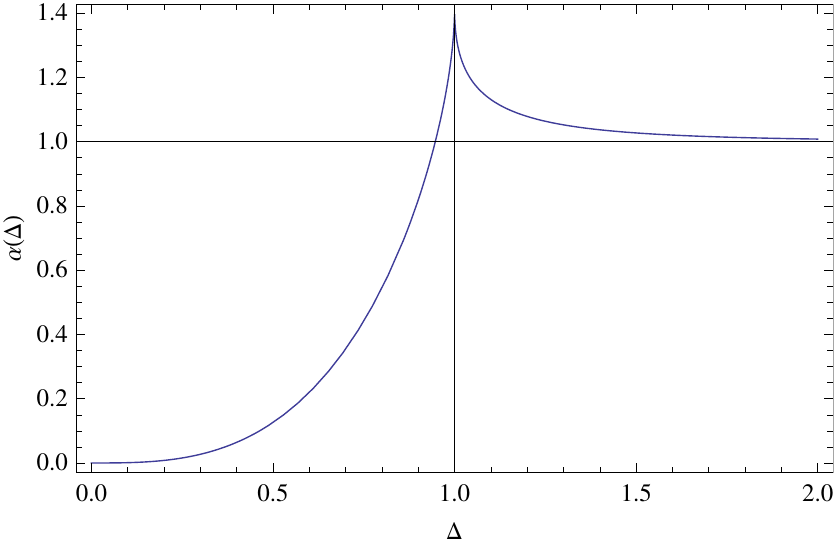}
\caption{\label{alphaDelta} The correction function $\alpha(\Delta)$ where $\Delta$ is taken relative to
$\Delta_*:=\sqrt{\gamma /2}\lP$. Note that $\alpha(\Delta)\rightarrow 1$ for $\Delta\gg\Delta*$.}
\end{minipage}
\hspace{0.5cm}
\begin{minipage}{0.5\linewidth}
\centering
\includegraphics[scale=0.75]{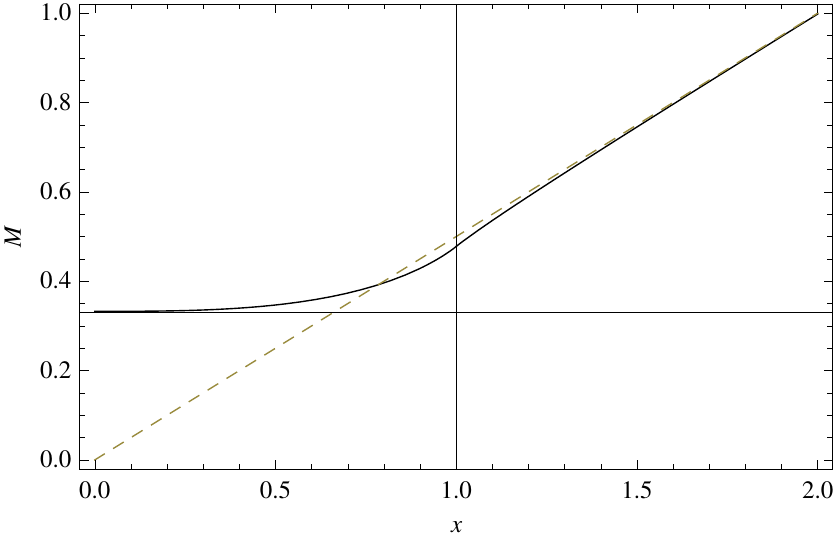}
\caption{\label{horizoncurve} Horizon curve: Right hand side of
$M=x/2f_{\alpha}(x)$ (solid) and the classical horizon curve (dashed).}
\end{minipage}
\end{figure}

In this case, in vacuum, one can solve the constraints and the equations of motion (in the static gauge $K_{x}=K_{\phi}=N^{x}=0$ and $E^{x}=x^{2}$) and obtain the quantum corrected version of the classical Schwarzschild solution (note that not all quantum gravity effects have been included)
\begin{equation} \label{Schw1}
{\rm d}s^{2}=-g_{\alpha}^{2}\left(1-\frac{2Mf_{\alpha}}{x}\right){\rm d}t^{2}+
\left(1-\frac{2Mf_{\alpha}}{x}\right)^{-1}{\rm d}x^{2}+x^{2}{\rm d}\Omega^{2}\,.
\end{equation}
Here $f_{\alpha}=1/g_{\alpha}$ are functions correcting the classical $E^{\phi}$ and $N$ and tend to one in the limit $x\rightarrow\infty$ giving back the classical solution. The horizon equation is given implicitly by $M=2x/f_{\alpha}$ and gives a mass threshold below which black hole does not form (see Fig.~\ref{horizoncurve}), consistent with the observation of \cite{Collapse}. Away from deep quantum regime one can repeat Hawking's analysis to find modified Hawking temperature $k_{\rm B}T=\hbar\alpha(x_{{\rm h}}) g_{\alpha}^{2}(x_{{\rm h}})/8\pi M$, which goes to the standard expression when the horizon $x_{h}\gg\sqrt{\gamma/2}l_{p}$. More importantly, since the constraint algebra is classical, spacetime covariance properties persist. This can be verified by doing a coordinate transformation to the Painlev\'e-Gullstrand coordinates,  in terms of which the metric becomes
\begin{equation} \label{corrected metric in painleve coordinates}
{\rm d}s^{2}=-\frac{1}{f_{\alpha}^{2}}\left(1-\frac{2Mf_{\alpha}}{x}\right){\rm d}T^{2}+
f_{\alpha}^{-2}{\rm d}x^{2}+2f_{\alpha}^{-2}\sqrt{f_{\alpha}^{2}-1+\frac{2Mf_{\alpha}}{x}}{\rm d}T{\rm
  d}x+ x^{2}{\rm d}\Omega^{2}\,,
\end{equation}
and checking that for this metric the constraints are identically satisfied (see \cite{ModSpacetime} for details).


\section*{Case II: $\bar{\alpha}=\alpha$} As already noted, in this case both the Hamiltonian and the constraint algebra get modified. Assuming that here also the usual spacetime properties continue to hold, we solve the constraints and the equations of motion to find the 'metric' corresponding to the classical Schwarzschild solution
\begin{equation}\label{Schwarz}
\not\!{{\rm d}}s^{2}=-\alpha^{-2}\left(1-\frac{2M}{x}\right)\not\!{\rm d}t^{2}+
\left(1-\frac{2M}{x}\right)^{-1}\not\!{\rm d}x^{2}+x^{2}\not\!{\rm
  d}\Omega^{2} \,. 
\end{equation}
Following the previous procedure if we now go to the corresponding 'Painlev\'e-Gullstrand' coordinates, the 'metric' becomes (slash on $ds$ and the quotes indicate that these quantities do not have the usual geometric meaning as made clear from statement after \eqref{PGds})
\begin{equation} \label{PGds}
\not\!{\rm d}s^{2}=-\alpha^{-2}\left(1-\frac{2M}{x}\right)\not\!{\rm d}T^{2}+
\alpha^{-2}\not\!{\rm d}x^{2}+2\alpha^{-2}\sqrt{\alpha^{2}-1+\frac{2M}{x}}
\not\!{\rm d}x\not\!{\rm d}T+x^{2}\not\!{\rm d}\Omega^{2}
\end{equation}
It turns out that \eqref{PGds} does not solve the constraints thus showing that for modified constraint algebra, coordinate transformations do not map solutions of constraints to other solutions.


\section{Implications and conclusions}
Having shown that the conventional spacetime properties (or even a concept like metric) do not hold for modified constraint algebra, the obvious next question is what happens to the concept of horizon? Do the entropy calculations in LQG \cite{ABCK:LoopEntro, KaulMaj, BHEntSU21}, based on classical concept of horizon, continue to hold? To explore such questions one can include matter (in the form of scalar field, say) as a second order perturbation to the classical solutions and evaluate various horizon conditions like the trapping horizon \cite{trapping} and the isolated horizon \cite{HorRev}. For classical spherically symmetric backgrounds (where classical algebra holds), these two notions of horizon are equivalent. However, for modified algebra the two notions of horizon turn out to be inequivalent (details omitted here for want of space can be found in \cite{ModSpacetime}).

 In fact the trapping horizon, given by $\ex/(E^\varphi)^2-\left(N^x/N\right)^2=0$, becomes gauge dependent giving different results in different 'coordinates' (defined through the choice of N, since as seen above, the concept of coordinates is not meaningful in this case). The trapping horizon condition can be modifed in a simple way to $\ex/(E^\varphi)^2-\left(N^x/\bar{\alpha}N\right)^2=0$ such that different coordinate systems give equivalent results (modification obtained looking at equations of motion and not based on some understanding of horizons in quantum gravity). 
With such a scenario, it seems that black hole entropy calculations, where the isolated horizon conditions are fixed before quantization, miss out some quantum features relevant for the horizon. This, for instance, can have bearing on the value of Barbero-Immirzi parameter. But more crucially it is to be noted that for corrected algebra even the concept of isolated horizon is difficult to define suggesting that a reconsideration of entropy calculations might be needed. On the other hand we also had an example where the quantum correction did not modify the algebra and where classical definitions continue to give consistent results and therefore entropy calculations in such a situation would also be consistent. However, here only one type of quantum gravity correction was considered. More generally there will be holonomy and quantum backreaction effects (which so far have been difficult to implement in full generality in inhomogeneous situations) and which most likely would lead to a modified algebra. In such a case one might need a quantum definition of the mapping from phase space to spacetime and even a quantum notion of classical concepts of geometry.

\section*{Acknowledgments}
The author would like to thank Martin Bojowald for suggesting improvements to the draft. This work is supported under Max Planck-India Partner Group in
Gravity and Cosmology.

\section*{References}


\begin{thebibliography}{31}

\bibitem{Rov}
Rovelli C 2004 Quantum Gravity \emph{Cambridge University Press, Cambridge, UK}

\bibitem{ThomasRev}
Thiemann T 2007 Introduction to Modern Canonical Quantum General Relativity \emph{Cambridge University Press, Cambridge, UK} [gr-qc/0110034]

\bibitem{ALRev}
Ashtekar A and Lewandowski J 2004 \emph{Class.\ Quantum Grav.} {\bf 21} R53--R152 [gr-qc/0404018]

\bibitem{AbSinginLQC}
Bojowald M 2001 \emph{Phys. Rev. Lett.} {\bf 86} 5227--5230 [gr-qc/0102069]

\bibitem{APS1}
Ashtekar A, Pawlowski T and Singh P 2006 \emph{Phys. Rev. Lett.} {\bf 96} 141301 [gr-qc/0602086]

\bibitem{ModSpacetime}
Bojowald M, Paily G, Reyes J and Tibrewala R 2011 \emph{Class. Quant. Grav.} {\bf 28} 185006 [arXiv:1105.1340] 

\bibitem{SphSymm}
Bojowald M 2004 \emph{Class.\ Quantum Grav.} {\bf 21} 3733--3753 [gr-qc/0407017]

\bibitem{SphSymmHam}
Bojowald M and Swiderski R 2006 \emph{Class.\ Quantum Grav.} {\bf 23} 2129--2154 [gr-qc/0511108]

\bibitem{Regained}
Hojman S, Kucha\v{r} K, and Teitelboim C 1976 \emph{Ann.\ Phys.\ (New York)} {\bf 96} 88--135

\bibitem{QSDV}
Thiemann T 1998 \emph{Class.\ Quantum Grav.} {\bf 15} 1281--1314 [gr-qc/9705019]

\bibitem{LTB}
Bojowald M, Harada T and Tibrewala R 2008 \emph{Phys.\ Rev.\ D} {\bf 78} 064057 [arXiv:0806.2593]

\bibitem{LTBII}
Bojowald M, Reyes J and Tibrewala R 2009 \emph{Phys.\ Rev.\ D} {\bf 80} 084002 [arXiv:0906.4767]

\bibitem{Collapse}
Bojowald M, Goswami R, Maartens R and Singh P 2005 \emph{Phys.\ Rev.\ Lett.} {\bf 95} 091302 [gr-qc/0503041]

\bibitem{ABCK:LoopEntro}
Ashtekar A, Baez J, Corichi A and Krasnov K 1998 \emph{Phys.\ Rev.\ Lett.} {\bf 80} 904--907 [gr-qc/9710007]

\bibitem{KaulMaj}
Kaul R and Majumdar P 1998 \emph{Phys.\ Lett.\ B} {\bf 439} 267--270 [gr-qc/9801080]

\bibitem{BHEntSU21}
Engle J, Noui K and Perez A 2010 \emph{Phys.\ Rev.\ Lett.} {\bf 105} 031302 [arXiv:0905.3168]

\bibitem{trapping}
Hayward S 1994 \emph{Phys.\ Rev.\ D} {\bf 49} 6467--6474

\bibitem{HorRev}
Ashtekar A and Krishnan B 2004
\emph{Living Rev.\ Rel.} {\bf 7} 1--77 [gr-qc/0407042]

\end{thebibliography}
\end{document}